\documentclass[preprintnumbers,amsmath,amssymb,nofootinbib]{revtex4}

\usepackage{lineno,hyperref,amsmath}
\usepackage{graphicx,subcaption,epstopdf}
\usepackage{color}

\begin{document}

\title{A compact accelerator for MHz high repetition rate soft x-ray free electron laser}

\author{Ji Qiang}
\address{Lawrence Berkeley National Laboratory,
          Berkeley, CA 94720}

	
	
\begin{abstract}

High-brightness X-ray Free Electron Lasers (FELs) produce spatially and temporally coherent pulses on attosecond to femtosecond timescales, providing a transformative tool for discovery across biology, chemistry, physics, and materials science. 
This paper 
proposes a compact accelerator that enables a high-repetition-rate (MHz) 
1 nm soft X-ray FEL with a footprint of less than 100 meters. Such an FEL
is suitable for installation within research institution settings where space is limited. The accelerator leverages a multi-turn recirculating linear accelerator that integrates state-of-the-art superconducting accelerator technology with recent advances in diffraction-limited storage rings. We present the conceptual layout and analyze the impact of two most challenging factors for such a compact accelerator, incoherent and coherent synchrotron radiation.
We have systematically studied both effects for different multi-bend achromat lattices
and electron beam peak currents. For a peak current of $60$ Ampere before final compression and using 11-bending 
magnets, the horizontal emittance growth after the 90-degree arc can be kept below
$10\%$,
demonstrating that these effects are not limiting factors for achieving high-quality electron beams.
Such a compact X-ray FEL facility would substantially reduce both construction and operational costs, greatly expanding access to these powerful research tools. 
Furthermore, this concept provides a potential upgrade path to generating hard X-ray radiation by incorporating high accelerating gradient structures to further accelerate a portion of the MHz electron beam.

\end{abstract}



\maketitle

\section{INTRODUCTION} 

X-ray light sources are indispensable tools for high-resolution experimental studies across a vast range of scientific disciplines, from structural biology and chemistry to materials science. The scientific grand challenges that next-generation photon sources must address have been well-defined, notably by the Basic Energy Sciences Advisory Committee in its report on ``Next Generation Photon Sources''~\cite{lanl}. These challenges demand the ability to probe two fundamental regimes: the temporal evolution of electronic, atomic, and chemical processes on femtosecond timescales, and the structural and spectroscopic imaging of nano-objects with nanometer spatial resolution.

X-ray Free Electron Lasers (FELs) are uniquely suited to meet these demands. Their attosecond-to-femtosecond pulse durations are ideal for studying ultrafast dynamics, while their spatial and temporal coherence enable imaging with unparalleled spatial and spectral resolution. Furthermore, FELs deliver brightness many orders of magnitude beyond that of conventional synchrotron radiation sources~\cite{workshop2009}. While emerging diffraction-limited storage rings also promise a significant increase in brightness over third-generation sources, they lack a critical capability for ultrafast science. Specifically, they do not possess temporal coherence, and their pulse durations of over 100 picoseconds are ill-suited for resolving the ultrafast dynamics central to many of today's scientific questions.

The past two decades have seen the successful construction and operation of several pioneering X-ray FELs, including LCLS, FERMI,  SACLA, PAL-XFEL, SwissFEL, SXFEL, FLASH and the European XFEL~\cite{lcls,fermi,jfel,palfel,swissfel,sxfel,flash,xfel}. While scientifically transformative,
the majority of these facilities
operate at a low repetition rate ($\sim100$ Hz), which severely limits the number of available user beamlines (typically fewer than five) and contributes to high operational costs per experiment.
To address these limitations, the global focus has shifted toward the development of high-repetition-rate (MHz) free electron lasers. This initiative will lead to the establishment of two new facilities including the soft/hard X-ray LCLS-II in the United States~\cite{lcls2,ding}
and a combined soft/hard X-ray FEL in China\cite{shine} in addition to
the hard X-ray XFEL in Germany. Despite their advanced capabilities, these machines require kilometer-scale footprints and investments exceeding one billion dollars. For instance, the recently commissioned LCLS-II operates at 4 GeV, with plans for an upgrade to 8 GeV to generate shorter wavelength X-rays, and spans over three kilometers in length. Similarly, other proposed next-generation MHz facilities, such as MariX and $S^3$FEL, are designed with substantial footprints of approximately 500 meters\cite{marix} and 1.7 km~\cite{s3fel} respectively. 
The FLASH soft X-ray FEL spans over 300 meters but offers radiation wavelengths above 3 nm.
The considerable scale and cost of these high-repetition-rate FELs highlight the need for a more compact and cost-effective concept.

The pursuit of compact X-ray FELs is an active area of research, with several concepts currently proposed or under construction. These concepts primarily rely on high-gradient, normal-conducting C-band or X-band RF structures to achieve a smaller footprint. However, this approach is fundamentally unsuited for the MHz-class repetition rates.
To the best of our knowledge, there is currently no X-ray FEL facility—existing or proposed—capable of generating 1 nm X-ray FEL radiation at a 1 MHz repetition rate within about a 100-meter footprint.
In this paper, we introduce a novel compact accelerator concept for a EUV to 1 nm soft X-ray FEL facility. This concept also includes the potential to divert a fraction of the MHz electron beam to drive a hard-X-ray FEL. Our approach utilizes superconducting radio-frequency (SRF) accelerating cavities—a technology capable of supporting MHz repetition rates—combined with a recirculating linear accelerator (linac) architecture. This allows us to achieve high throughput within a dramatically reduced facility footprint.

Recirculating linear accelerators have been proposed for X-ray FELs in previous studies~\cite{lux,ukrec,york,marix2}. Compared to those concepts, the compact accelerator for XFEL facility presented in this paper features a significantly smaller footprint and employs several novel techniques: (1) folding the low-emittance electron beam injector into the transverse dimension to reduce the total footprint; (2) using two superconducting linacs positioned on opposite sides of the recirculating linac for three-pass acceleration of the electron beam to nearly 2 GeV; (3) replacing the traditional 180-degree return arcs with two 90-degree arcs separated by straight transfer line sections dedicated to beam diagnostics, manipulation, heating, seeding, and path length control; (4) adopting the compact multi-bend achromat (MBA) design from diffraction-limited synchrotron light sources to preserve small transverse emittances; (5) utilizing 90-degree arcs for initial modest bunch compression to maintain electron beam quality, followed by a large bunch compressor for final compression to achieve kilo-Ampere peak currents; and (6) employing a passive dechirper after final compression to reduce the correlated energy spread of the electron beam. Together, these techniques enable the generation of the high-brightness electron beam required for the proposed compact XFEL facility.

In the above compact recirculating linac, 
two challenging effects, 
Incoherent Synchrotron Radiation (ISR)
and Coherent Synchrotron Radiation (CSR), have to be addressed.
To minimize the longitudinal footprint of the linac, the height of the 90-degree arc is reduced, resulting in a small bending radius. For a GeV-level electron beam passing through the 90-degree arc with such a radius, ISR and CSR effects become dominant and can significantly degrade beam quality if not properly handled. In this study, we systematically analyze these effects using both analytical estimates and detailed numerical simulations.

 The organization of this paper is as follows: Following the Introduction, we present the compact accelerator concept for high-repetition-rate X-ray FEL facility in Section 2, study emittance and energy spread growth from the ISR and CSR effects in Section 3, and discuss potential challenges and future work in Section 4.

\section{A compact accelerator for high repetition rate soft X-ray FEL facility}

Current X-ray FEL facilities worldwide utilize straight RF linear accelerators to generate high-brightness electron beams, which subsequently pass through an undulator to produce coherent X-ray radiation. In this paper, we propose a novel compact recirculating RF accelerator as an alternative to the conventional straight linear accelerator, which typically spans several hundred to thousands of meters. This concept significantly reduces the facility's footprint. Additionally, the recirculating accelerator decreases the number of required RF cavities and enhances both RF and cooling efficiency. Consequently, the compact X-ray FEL offers a more cost-effective solution.
Furthermore, to achieve compactness in this new concept, we propose dividing each 180-degree arc into two 90-degree arcs separated by a straight section. These straight sections accommodate beam diagnostics and manipulation, laser heaters, and even external seeding. The photoinjector is also integrated into a straight section to further minimize the accelerator's length. Compact multi-bend achromats are employed for the 90-degree arcs. Beyond compactness, the recirculating architecture mitigates the transport of dark current into the undulator, offering an advantage over straight linear accelerators.
 
\begin{figure}[!htb]
\centering
\includegraphics[width=10. cm]{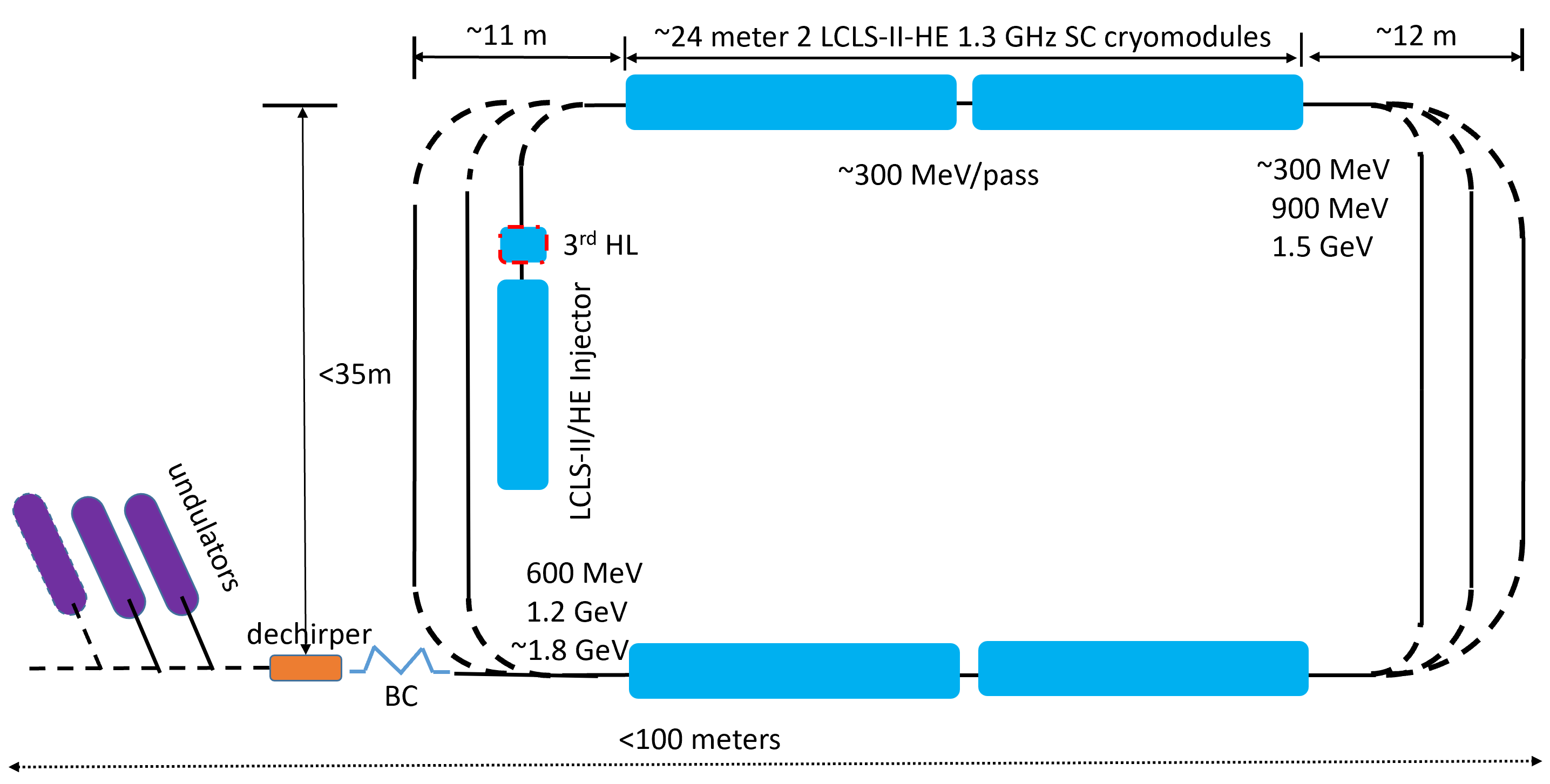}
\caption{
Schematic plot of the compact x-ray FEL facility.}
\label{figlayout}
\end{figure}  

A layout of the compact accelerator for X-ray FEL is shown in Fig.~\ref{figlayout}. 
It consists of a high-repetition-rate LCLS-II/HE-type injector, a 3.9 GHz third-harmonic linearizer, a 90-degree achromatic arc, a 4-meter matching section, and two cryomodules comprising 16 1.3 GHz superconducting cavities designed for LCLS-II/HE~\cite{lcls2he}. The layout further includes a second 4-meter matching section, three additional 90-degree achromatic arcs, three straight transfer lines, another set of three 90-degree achromatic arcs, a third 4-meter matching section, two more superconducting cavity cryomodules, another matching section, two additional 90-degree achromatic arcs, two straight transfer lines, another pair of 90-degree achromatic arcs, a bunch compressor (BC), a passive longitudinal dechirper, switch kickers, and a farm of FEL undulators. 
The longitudinal height of the outermost arc on the right side is approximately 8 meters, while the longitudinal height of the outermost arc on the left side is approximately 7 meters. 
The length of each of the two superconducting cryomodules is about 24 meters, based on the LCLS-II design. 
This results in a total length for the electron accelerator of less than 50 meters.

A 100 pC electron beam from the high-repetition-rate injector exhibits a peak current of approximately 10 A~\cite{lcls2inj}. A third-harmonic superconducting cavity module is employed to control or linearize the longitudinal phase space downstream. After this module, the electron beam achieves an energy of about 50 to 80 MeV. This beam is then transported through the first top-left 90-degree achromatic arc into the top superconducting linac section. The maximum accelerating gradient of the 1.3 GHz superconducting cavity is assumed to be 20 MV/m, as utilized in the LCLS-II HE project.

After acceleration in the top superconducting linac, the electron beam reaches an energy of about 300 MeV. It is subsequently transported through the first top-right 90-degree achromatic arc into a straight transfer line, where beam diagnostics and a laser heater can be installed. Following this transfer line, the beam passes through the first bottom-right 90-degree achromatic arc and enters the bottom superconducting linac section, which employees the same type of RF cavities as the top linac. After acceleration in this linac, the beam attains an energy of approximately 600 MeV.

The beam is then transported through a series of arcs and transfer lines, including the first bottom-left 90-degree achromatic arc, the first left straight transfer line, and the second top-left 90-degree achromatic arc, before re-entering the top superconducting linac. After this second pass, the beam reaches an energy of about 900 MeV. It is subsequently transported through the second top-right 90-degree achromatic arc, another straight transfer line, and the second bottom-right 90-degree achromatic arc before entering the bottom superconducting linac for the second time, achieving an energy of 1.2 GeV.

This process continues with the beam passing through the second bottom-left 90-degree achromatic arc, the second left straight transfer line, and the third top-left 90-degree achromatic arc to enter the top linac for the third time. After this acceleration, the beam attains an energy of approximately 1.5 GeV. It is then transported through the third top-right 90-degree achromatic arc, the third right transfer line, and the third bottom-right 90-degree achromatic arc before entering the bottom superconducting linac for the third time, reaching a final energy of about 1.8 GeV.

Finally, the electron beam is extracted from the recirculating superconducting linac and undergoes final compression in a bunch compressor, achieving a peak current of approximately 1 kA. The chirp in the electron beam after compression is removed using a passive dechirper~\cite{dechirp0,dechirp1,dechirp2} before the beam is directed into X-ray FEL undulators to produce coherent X-ray radiation.

The performance of the X-ray FEL radiation can be estimated using the electron beam from the compact recirculating linac described above and the analytical model in reference~\cite{huang}. The resonant equation for X-ray FEL coherent radiation is given by:
\begin{equation}
    \lambda_r = \frac{\lambda_u}{2\gamma^2}\left(1 + \frac{K_0^2}{2}\right)
\end{equation}
where $\lambda_r$ is the radiation wavelength, $\lambda_u$ is the undulator period, $\gamma$ is the relativistic factor of the electron beam, and $K_0$ is the undulator parameter. For a 1 nm soft X-ray radiation with an undulator period of 15 mm, the undulator parameter is calculated to be 1.14.

An important parameter influencing X-ray FEL radiation performance is the FEL parameter $\rho$, defined as:
\begin{equation}
    \rho = \left[\frac{1}{16} \frac{I_e}{I_A} \frac{K_0^2 [JJ]^2}{\gamma^3 \sigma_x^2 k_u^2}\right]^{1/3}  
\end{equation}
where $I_e$ is the electron peak current, $I_A \approx 17$ kA is the Alfvén current, $[JJ] = J_0(\xi) - J_1(\xi)$ with $J_0$ and $J_1$ being the 0th- and 1st-order Bessel functions, respectively, and $\xi = K_0^2/(4 + 2K_0^2)$, and $k_u = 2\pi/\lambda_u$ is the wave number associated with the undulator period. Assuming a final electron beam normalized emittance of 0.5 mm mrad and an average Twiss beta function of 10 m, the FEL parameter is $\rho = 7 \times 10^{-4}$. For emittances of 1 mm mrad and 2 mm mrad, the values are $\rho = 5.6 \times 10^{-4}$ and $\rho = 4.4 \times 10^{-4}$, respectively.

The saturation power of the X-ray FEL radiation depends on the FEL parameter and is approximated by:
\begin{equation}
    P_{sat} \approx \rho \frac{E_e I_e}{e}
\end{equation}
For the aforementioned FEL parameters, the corresponding saturation powers are approximately 1.3 GW, 1.0 GW, and 0.8 GW for the three emittance values, respectively.

The X-ray FEL power gain length is given by:
\begin{equation}
    L_{G0} = \frac{\lambda_u}{4\pi \sqrt{3} \rho}
\end{equation}
For the three final emittance values, the corresponding gain lengths are 0.99 m, 1.23 m, and 1.57 m, respectively. Typically, it requires approximately 20 gain lengths for the coherent X-ray radiation to reach saturation, resulting in undulator lengths of 20 m, 24 m, and 32 m, respectively. Accounting for gaps between undulator sections, the maximum undulator hall length is expected to be less than 50 m.

Combining the undulator hall length with the less than 50 m length of the recirculating superconducting linac, the total length of the X-ray FEL facility is less than 100 m. This compact footprint makes the facility suitable for installation in space-constrained universities or research institutions.

To produce effective coherent X-ray radiation in the FEL, the final electron beam emittance should satisfy the following condition:
\begin{equation}
    \epsilon_n < \gamma \frac{\lambda_r}{4 \pi} \frac{\bar{\beta}}{L_{G0}}
\end{equation}
Using the gain lengths and the average Twiss beta function value discussed above, this results in $\epsilon_n < 2.8$ mm mrad, $\epsilon_n < 2.3$ mm mrad, and $\epsilon_n < 1.8$ mm mrad for the respective cases. The first two conditions are clearly satisfied for the assumed final emittances of 0.5 mm mrad and 1 mm mrad. However, the assumed emittance of 2 mm mrad is near the boundary. This suggests that, for effective X-ray FEL radiation, the final electron beam emittance from the recirculating linac should be less than 2 mm mrad.

In the preceding discussion, we have demonstrated that an electron beam with an energy of 1.8 GeV, a peak current of 1 kA, and a transverse emittance of less than 2 mm mrad can produce GW-level coherent X-ray radiation at 1 nm within a 50 m undulator hall. In the following sections, we discuss how to achieve these final electron beam parameters.

We begin with considerations of longitudinal beam dynamics. A final compression factor of 10 to 20 is assumed, which must be achieved through over-compression due to the positive energy chirp of the electron beam before the final bunch compressor. This positive energy chirp is necessary to compress the electron beam using the 90-degree achromatic arcs, which have negative $R_{56}$ values. For the same absolute compression value but with over-compression, the final relative energy chirp $h_f = \frac{d(\Delta \gamma_f / \gamma_f)}{dz_f}$ of the electron beam before compression is given by:
\begin{equation}
    h_f = -\frac{1}{R_{56f}} \left(1 + \frac{1}{|C_f|}\right)
\end{equation}
where $R_{56f}$ is the momentum compaction factor of the last bunch compressor, and $C_f$ is the final bunch compression factor. Assuming $R_{56f} = -0.2$ m and a final compression factor of 10 to 20, the final relative energy chirp should range from 5.25 m$^{-1}$ to 5.5 m$^{-1}$.

For the proposed recirculating linac shown in Fig.~\ref{figlayout}, the final energy deviation can be expressed as:
\begin{equation}
    \Delta \gamma_f = \left( h_0 \gamma_0 \prod_{i=1}^6 C_i + \bar{V}_1 k \sin{\phi_1} \sum_{j=1}^3 \prod_{i=2j}^6 C_i + \bar{V}_2 k \sin{\phi_2} \sum_{j=1}^2 \prod_{i=2j+1}^6 C_i + \bar{V}_2 k \sin{\phi_2} \right) z_f
\end{equation}
where $h_0$ is the initial relative energy chirp after the third-harmonic linearizer, $C_i$ is the compression factor after arc $i$ (with the first arc being the top-left 90-degree achromatic arc after the linearizer), $\bar{V}_1$ and $\bar{V}_2$ are the normalized voltages of the top and bottom superconducting linacs, respectively, $\phi_1$ and $\phi_2$ are the design phases of the top and bottom linacs, respectively, and $k$ is the wave number corresponding to the RF frequency.

Assuming both top and bottom linacs operate at the same voltage $\bar{V}$ and phase $\phi$, all arcs except the first have the same compression factor $C$, and neglecting the initial energy chirp, the final energy deviation simplifies to:
\begin{equation}
    \Delta \gamma_f = \bar{V} k \sin{\phi} \frac{C^6 - 1}{C - 1} z_f
\end{equation}
The final relativistic factor $\gamma_f$ after three acceleration passes, neglecting the initial energy, is:
\begin{equation}
    \gamma_f = 6 \bar{V} \cos{\phi}
\end{equation}
The resulting final relative energy chirp is then:
\begin{equation}
    h_f = k \tan{\phi} \frac{C^6 - 1}{6 (C - 1)}
    \label{chirp}
\end{equation}

In this proposed facility, the final peak current is approximately 1 kA. Given a compression factor of 10 to 20 in the last bunch compressor, the electron beam peak current before final compression should be between 50 A and 100 A. With an initial peak current of 10 A from the injector, the total compression factor through the five major arcs (including straight sections) should be between 5 and 10. This implies a compression factor per arc of 1.38 to 1.59. 
Using Eq.~\ref{chirp}, this corresponds to an RF phase of approximately 2.70 to 4.25 degrees for the superconducting linac to achieve a chirp of 5.25 m$^{-1}$ to 5.5 m$^{-1}$.
These RF design phases have minimal impact on the final achievable electron beam energy.

\section{Effects of ISR and CSR in arcs}
\begin{figure}[!htb]
\centering
\includegraphics[width=11cm, height=3cm]{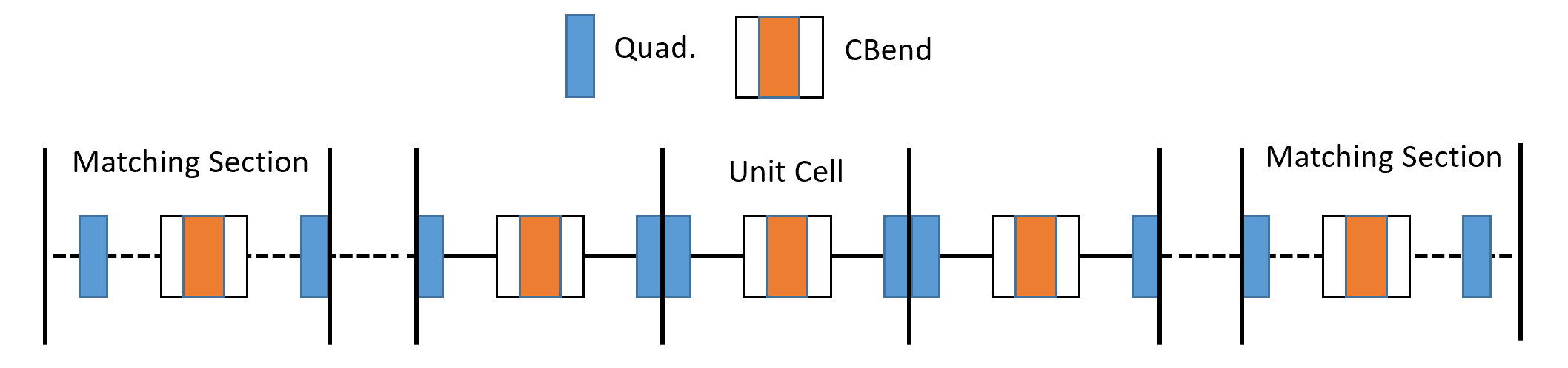}
\caption{Schematic layout of a multi-bend achromat 90-degree arc.}
\label{figmba}
\end{figure}

In the recirculating superconducting electron linac described above, multiple 90-degree achromatic arcs are employed. Leveraging recent advances in diffraction-limited synchrotron light source storage ring design, in this study, we propose to use a multi-bend achromat (MBA) system to minimize emittance growth~\cite{mba}. 
A schematic plot of the MBA system is shown in Fig.~\ref{figmba}.

This MBA lattice consists of a number of periodic cells, each comprising a combined-function bending magnet, two drift spaces, and two half-quadrupoles, along with two matching cells on either side of the periodic cells. In this study, the two matching cells are assumed to be identical, each containing a combined-function bending magnet, a drift space, and a quadrupole. It is further assumed that the combined-function bending magnet in the matching cell is identical to that in the periodic cells, and the quadrupole in the matching cell has the same strength as the quadrupoles in the periodic lattice.

To achieve an achromatic multi-bend arc, the drift length $L_d$ between the bending magnet and quadrupole in the matching cell, as well as the quadrupole length $L_q$, can be determined as follows. Assuming the linear transfer matrix through the periodic cells is denoted by $R$, the periodic dispersion and its derivative at the end of the cells can be obtained by solving the following equation:

\begin{equation}
\left(\begin{array}{c}
D \\
D' \\
1 
\end{array} \right) =   \left( \begin{array}{ccc}
		R_{11} & R_{12} & R_{16} \\
		R_{21} & R_{22} & R_{26} \\
        0 & 0 & 1 \\
	\end{array} \right)
\left(\begin{array}{c}
D \\
D' \\
1 
\end{array} \right)
\end{equation}
The solution of the above equation results in:
\begin{equation}
\left(\begin{array}{c}
D \\
D' 
\end{array} \right) =   \frac{1}{2-R_{11}-R_{22}}
\left(\begin{array}{c}
(1-R_{22})R_{16}+R_{12}R_{26}\\
R_{21}R_{16}+(1-R_{11})R_{26}
\end{array} \right)
\end{equation}
These dispersions are set to zero outside two matching cells to form an achromat arc by selecting the appropriate drift length and quadrupole length in the matching cells as follows:
\begin{eqnarray}
    L_q &=& \frac{1}{\sqrt{K_q}} \arcsin{\left( \frac{R^b_{26}}{D \sqrt{K_q}} \right)}, \\
    L_d &=& -\frac{R^b_{16}}{R^b_{26}} - \frac{R^q_{22}}{R^q_{21}},
\end{eqnarray}
where $R^q_{21}$ and $R^q_{22}$ are the linear transfer matrix elements of the quadrupole, and $R^b_{16}$ and $R^b_{26}$ are the linear transfer matrix elements of the combined-function bending magnet.

\begin{figure}[!htb]
\centering
\includegraphics[width=6.0cm]{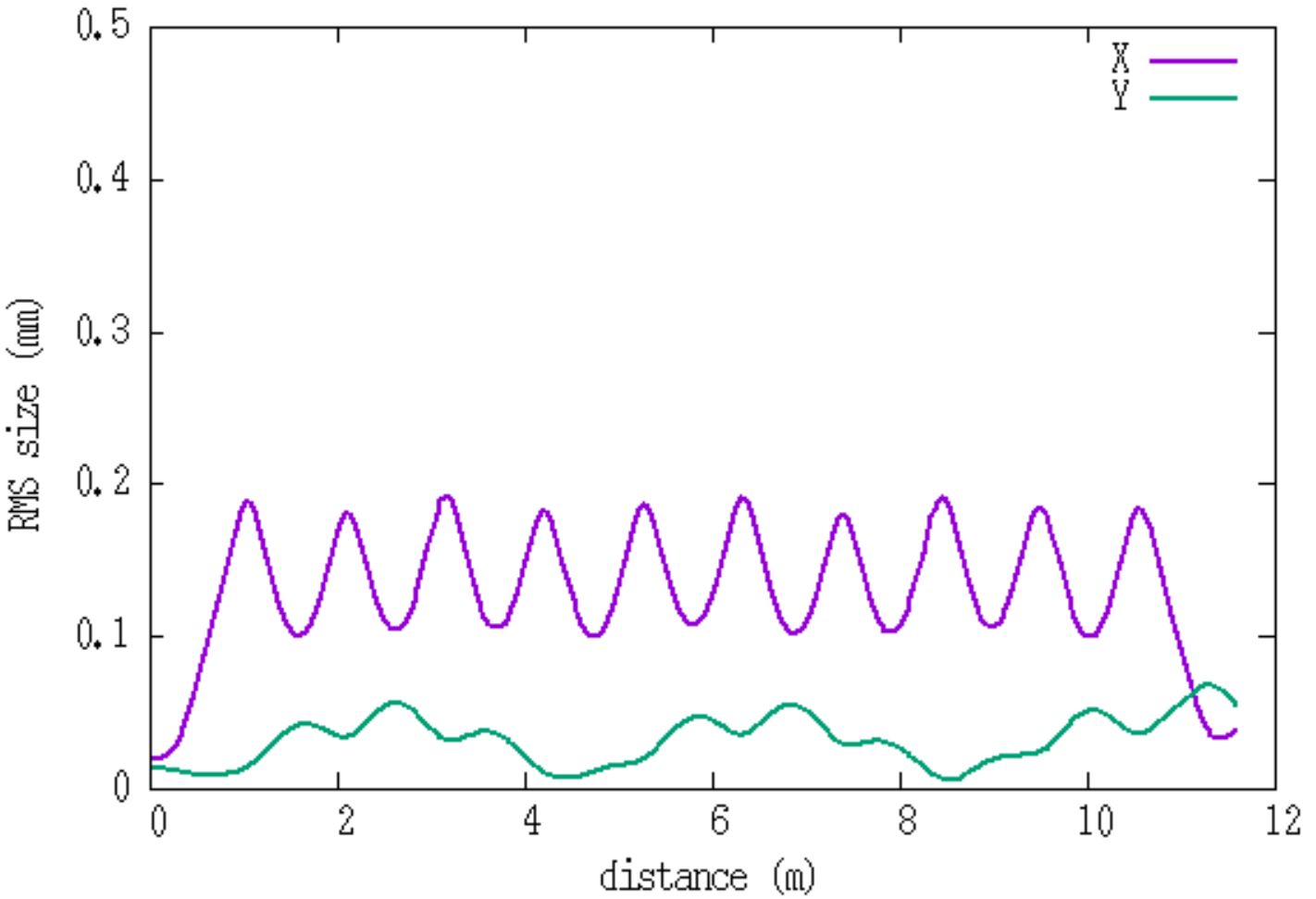}
\includegraphics[width=6.0cm]{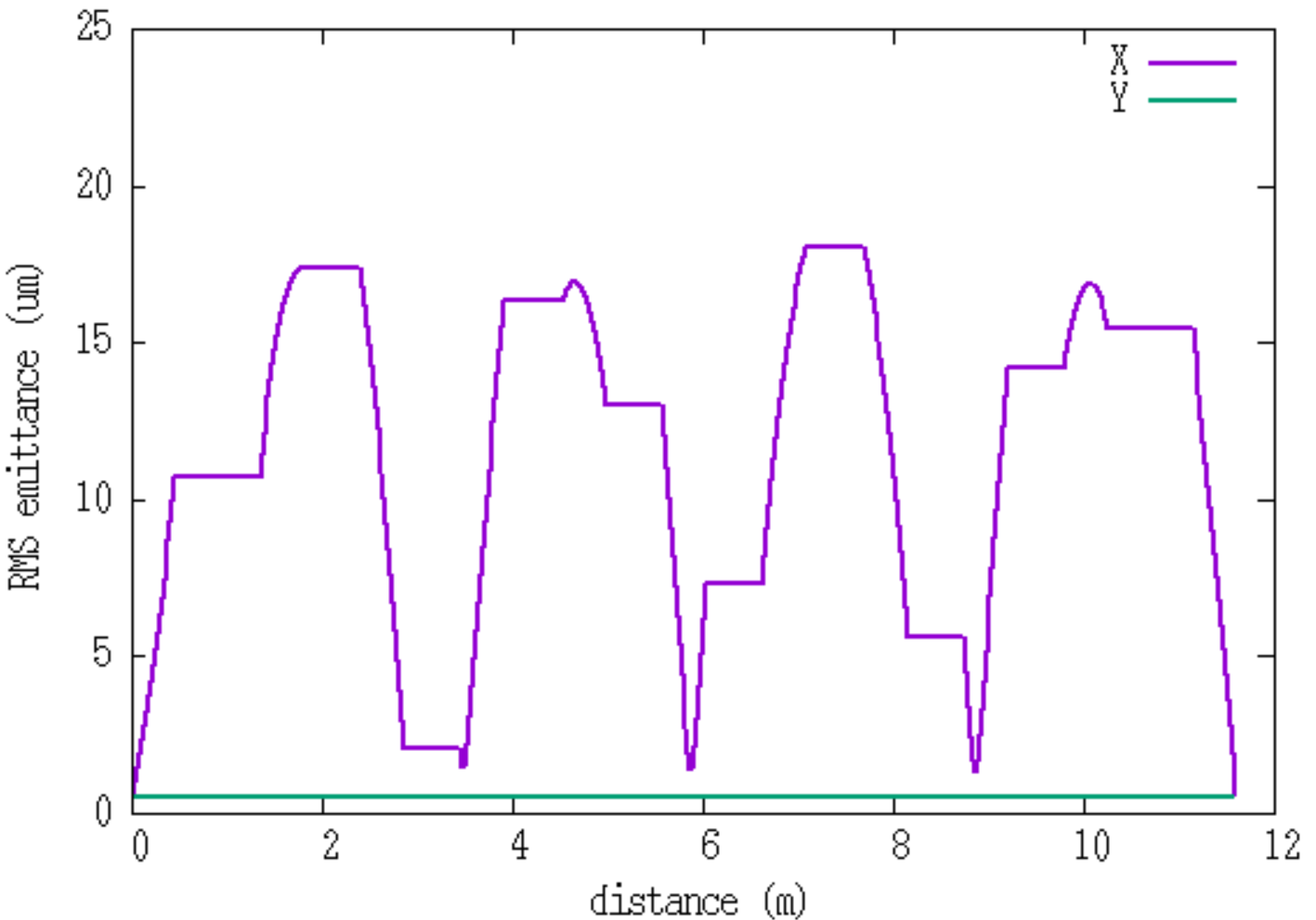}
\caption{Electron beam RMS size evolution (left) and emittance evolution (right) through a 90-degree, 11-bend achromat arc with a beam energy of 1.2~GeV and zero current.}
\label{figrms0a}
\end{figure}

As an illustration, Fig.~\ref{figrms0a} shows the evolution of the Root Mean Square (RMS) beam size and emittance of an electron beam through an 11-bend, 90-degree achromat arc. The dipole field within the 0.457-meter-long bending magnet is 1.25~T. Such a type of
bending magnet has been attained in the synchrotron light source community~\cite{bec}. In the periodic cell, the length of the focusing quadrupole is 0.2~m, and the separation between the quadrupole and the combined-function bend is also 0.2~m. The height of this arc is approximately 7~m, and the width is approximately 8~m.

In this example, the electron beam energy is 1.2~GeV. No collective effects are included in the simulation; only a linear transfer map was used for tracking. The initial electron beam distribution is transversely matched to the periodic cells and has an initial longitudinal relative energy chirp of 2~m$^{-1}$. The horizontal phase advance per cell within the periodic cells is 69~degrees, and the vertical phase advance is 47~degrees. There is a slight mismatch throughout the entire lattice. The defocusing quadrupole within the combined-function bend has a strength of $K = -5/m^2$, while the focusing quadrupole has a strength of $k = 12.65/m^2$. In the matching cells, the drift length is $L_d = 0.48$~m, and the quadrupole length is $L_q = 0.11$~m.

The horizontal RMS size is dominated by regular dispersion oscillations, which exhibits a minimum inside the bending magnet. The vertical RMS size shows some mismatch. However, with the linear transfer map, the emittance after the arc returns to its initial value of 0.5~$\mu$m. There is no emittance growth due to dispersion, thanks to the achromat design.

For a compact recirculating linac at GeV level, 
two most challenging effects are the 
Incoherent Synchrotron Radiation (ISR)
and Coherent Synchrotron Radiation (CSR).
The compact arc of the linac suggest the need for small bending radius in order to minimize the height of
the arc. Under such a condition, the ISR
and CSR effects become dominant and can
cause significant increase of the electron beam uncorrelated energy spread and transverse
emittance. In the following, we will address
these effects using both analytical estimates
and detailed numerical simulations.

ISR contributes to the growth of uncorrelated energy spread. An analytical estimate of the energy spread increase in eV through a 180-degree arc is given by~\cite{doug}:
\begin{equation}
    \sigma_E = \frac{\sqrt{0.118 \gamma^7}}{\rho} \times 10^{-7},
\end{equation}
where $\rho$ is the bending radius. For a 1.5~GeV electron beam passing through an arc with a 4-meter bending radius under a 1.25~T magnetic field strength within the bend, this results in an energy spread of approximately 12~keV. For a 1.2~GeV electron beam passing through the same magnetic field arc, the energy spread is approximately 7~keV. This increase in energy spread due to ISR was also verified using the Fokker-Planck model in reference~\cite{isr}. We measured the final uncorrelated energy spread growth through an 11-bend, 90-degree arc at 1.2~GeV as approximately 5~keV and at 1.5~GeV as approximately 9~keV. The simulation results agree well with the analytical estimates after taking into 
account the contribution from another 90-degree arc. The total ISR contribution to the uncorrelated energy spread across eleven arcs is expected to be less than 20~keV.

The normalized emittance growth due to quantum excitation from ISR through a 180-degree arc can also be estimated as~\cite{doug}:
\begin{equation}
    \Delta \epsilon_n = 7.19 \times 10^{-28} \frac{\gamma^6}{\rho^2} \langle H \rangle,
\end{equation}
where $\langle H \rangle$ is related to the Twiss beta function and dispersion inside the bend. For a 1.5~GeV beam passing through an arc with a 4-meter bending radius and $\langle H \rangle = 1$~m, this results in a normalized emittance growth of approximately 0.03~$\mu$m. For a 1.2~GeV electron beam through an arc with a 3.2~m bending radius, the ISR-induced emittance growth is approximately 0.01~$\mu$m. This suggests that the ISR effects in the proposed multi-bend achromat arcs are negligible. This conclusion is also supported by the macroparticle simulations discussed below.

The effects of Coherent Synchrotron Radiation (CSR) on normalized beam emittance growth through a bending magnet can be estimated by including contributions from both the longitudinal and transverse components of the CSR field as~\cite{dimitri2020}:
\begin{eqnarray}
    \Delta \epsilon_{n,L} &=& 7.5 \times 10^{-3} \frac{\beta_T}{\gamma} \left( \frac{r_e N L_B^2}{\rho^{5/3} \sigma_z^{4/3}} \right)^2, \\
    \Delta \epsilon_{n,T} &=& 2.5 \times 10^{-2} \frac{\beta_T}{\gamma} \left( \frac{r_e N L_B}{\rho \sigma_z} \right)^2,
\end{eqnarray}
where $\beta_T$ is the average bend-plane betatron function in the dipole magnet, $\gamma$ is the relativistic Lorentz factor for the beam's mean energy, $r_e$ is the classical electron radius, $N$ is the number of electrons in the beam, $L_B$ is the length of the bending magnet, $\rho$ is the bending radius of the magnet, and $\sigma_z$ is the RMS bunch length. For an electron beam with a charge of approximately 100~pC and an RMS bunch length of approximately 0.7~mm (corresponding to a peak current of about 10~A), using a 1.25~T dipole magnet and assuming $\beta_T = 10$~m inside the dipole magnet, the normalized emittance growth through a 90-degree arc is approximately 0.0004~$\mu$m with 11 dipoles, 0.0007~$\mu$m with 9 dipoles, and 0.001~$\mu$m with 7 dipoles for a 1.2~GeV electron beam. For a 1.5~GeV electron beam, the growth is approximately 0.001~$\mu$m with 7 dipoles, 0.0006~$\mu$m with 9 dipoles, and 0.0004~$\mu$m with 11 dipoles. Using more dipole magnets for a 90-degree arc results in reduced emittance growth. 

Further increasing the electron beam peak current by reducing the RMS bunch length by a factor of 10 (corresponding to a peak current of approximately 100~A) results in significantly larger CSR-induced emittance growth. For a 1.2~GeV electron beam, the normalized emittance growth after a 7-dipole, 90-degree arc is approximately 0.5~$\mu$m, about 0.2~$\mu$m after a 9-dipole arc, and about 0.1~$\mu$m after an 11-dipole, 90-degree arc. The results are comparable for a 1.5~GeV electron beam. While these emittance growth values are significant, they are not unacceptable. This suggests that the electron beam peak current should be kept below 100~A and that more bending magnets should be used to mitigate CSR-induced emittance growth.

\begin{figure}[!htb]
\centering
\includegraphics[width=6.0cm,height=4.5cm]{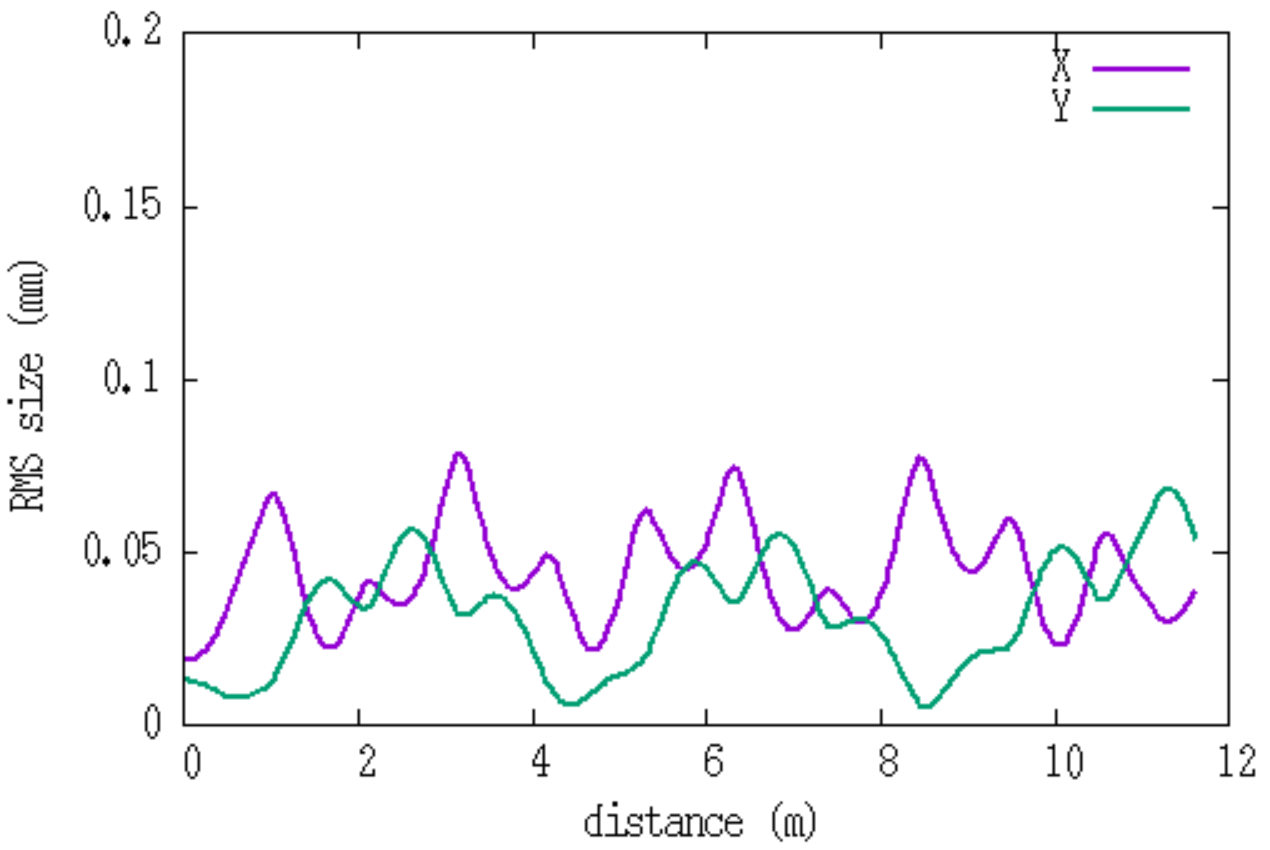}
\includegraphics[width=6.0cm,height=4.5cm]{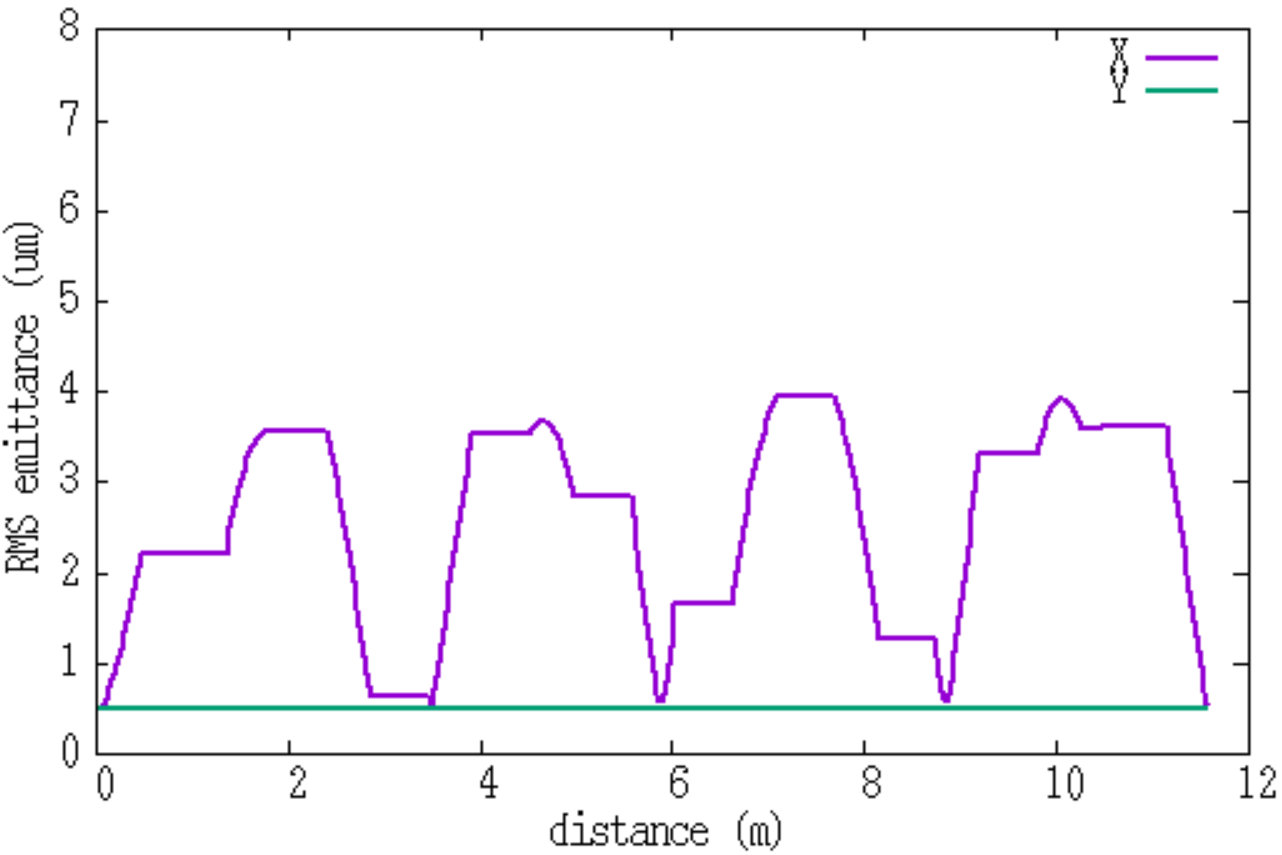}
\caption{Electron beam RMS size evolution (left) and emittance evolution (right) through a 90-degree, 11-bend achromat arc with a beam energy of 1.2~GeV and an initial peak current of 60~A.}
\label{figrms100a}
\end{figure}

Next, we evaluate the effects of element nonlinearity by using an exact model for drift, incorporating energy dependence in the quadrupole, employing a fifth-order transfer map for the bending magnet, and considering the effects of ISR,
one-dimensional CSR,
and space-charge using particle tracking. These evaluations are conducted for the same electron beam as described above, but with an initial peak current of 60~A. Figure~\ref{figrms100a} shows the RMS size and emittance evolution through the arc, obtained from macroparticle simulations that include all the aforementioned effects, performed using the parallel beam dynamics code IMPACT~\cite{impact1,impact2}.

With the 60 A peak current, the transverse sizes are mismatched. The emittance evolution exhibit patterns similar to those observed in the zero-current case discussed earlier. However, there is a slight final horizontal emittance growth of approximately 5\%, with the dominant contribution of about 3\% arising from lattice element nonlinearity.

\begin{figure}[!htb]
\centering
\includegraphics[width=6.0cm,height=4.5cm]{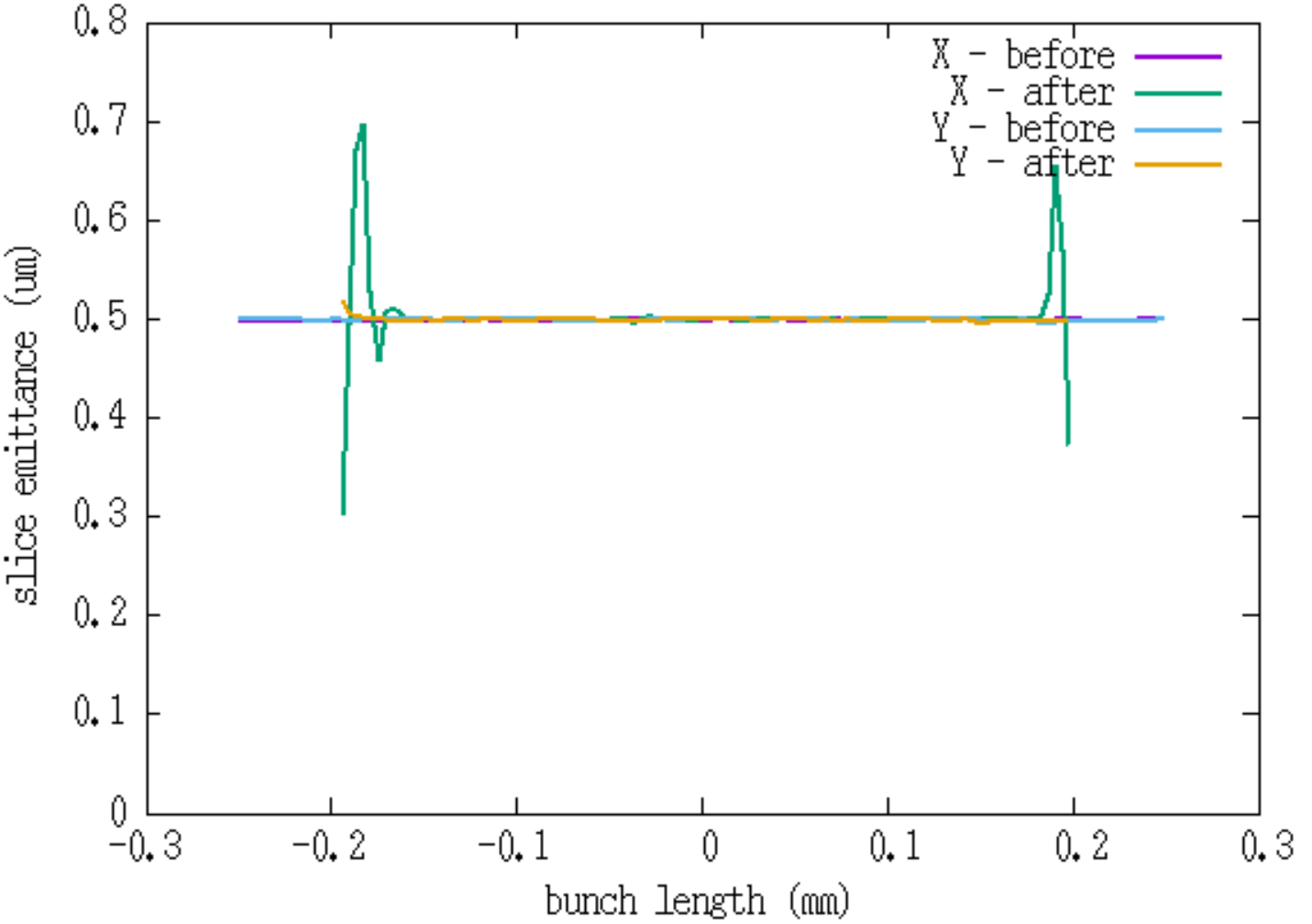}
\caption{Electron beam transverse slice emittance before and after the 90-degree, 11-bend achromat arc with a beam energy of 1.2~GeV and an initial peak current of 60~A.}
\label{figslemt}
\end{figure}
Figure~\ref{figslemt} illustrates the slice emittances before and after the 90-degree arc. There is minimal emittance growth around the core of the electron beam. The significant growth observed at both edges is due to the sharp-edge assumption of the flat-top current profile.

\begin{figure}[!htb]
\centering
\includegraphics[width=6.0cm,height=4.5cm]{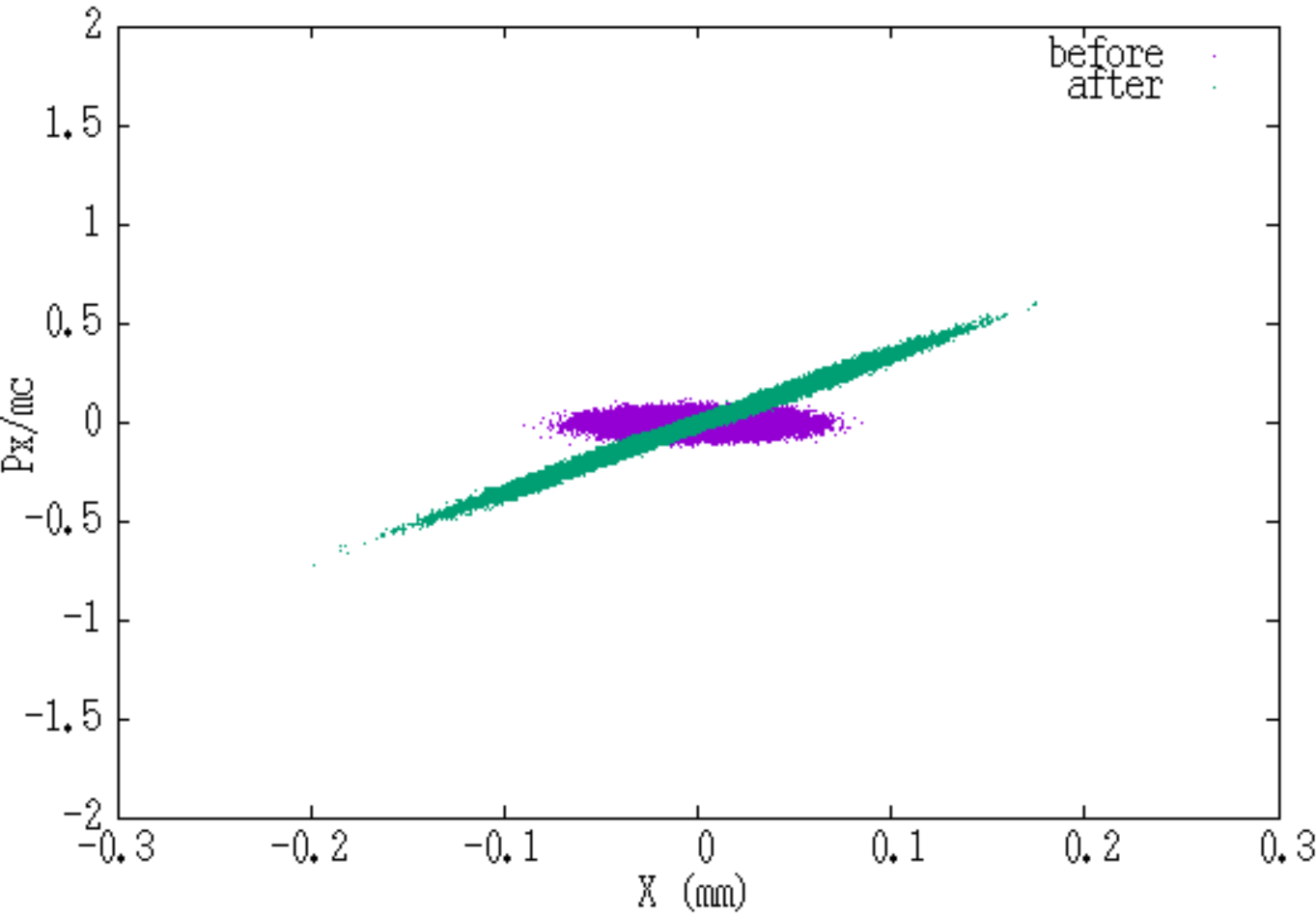}
\includegraphics[width=6.0cm,height=4.5cm]{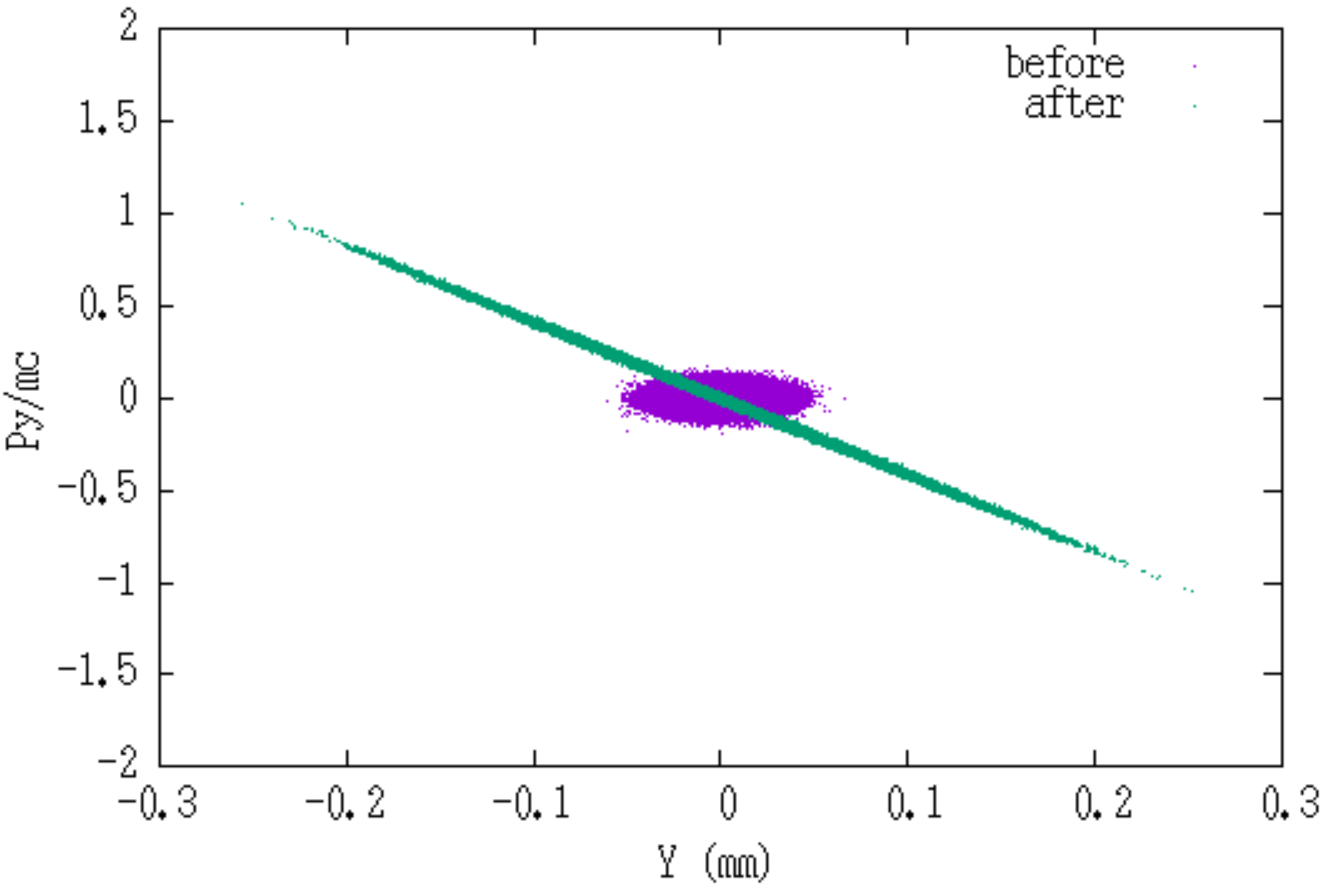}
\caption{Electron beam transverse phase space particle distribution before and after the 90-degree, 11-bend achromat arc with a beam energy of 1.2~GeV and an initial peak current of 60~A.}
\label{figtrphs}
\end{figure}
Figure~\ref{figtrphs} shows the electron beam
transverse phase space particle distribution before and after 
the 90-degree arc. There is no noticeable distortion in the transverse phase space due to
the nonlinear effects through the arc.

\begin{figure}[!htb]
\centering
\includegraphics[width=6.0cm,height=4.5cm]{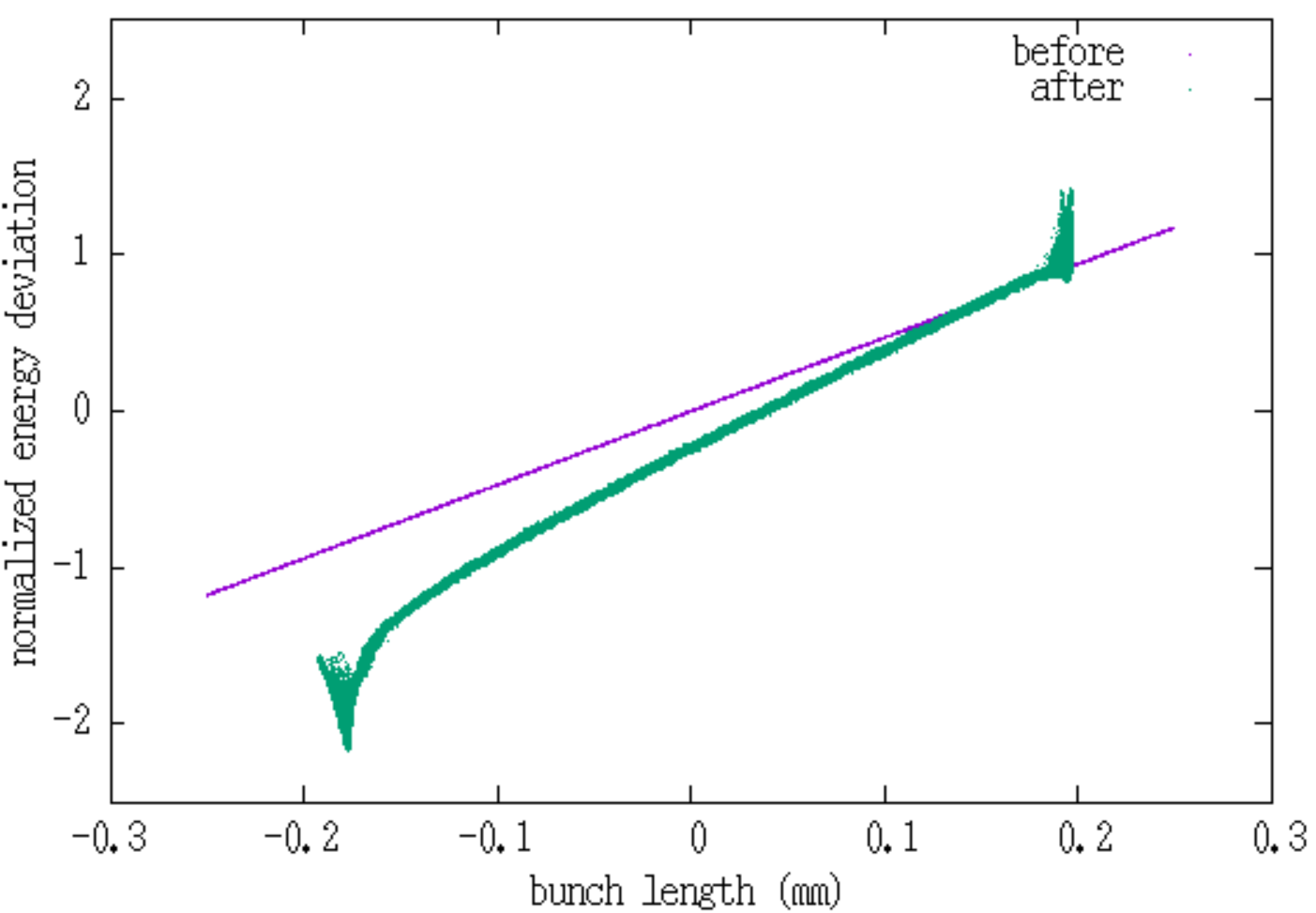}
\caption{Electron beam longitudinal phase space particle distribution before and after the 90-degree, 11-bend achromat arc with a beam energy of 1.2~GeV and an initial peak current of 60~A.}
\label{figlphs}
\end{figure}
Figure~\ref{figlphs} displays the longitudinal phase space particle distribution. The majority of the beam maintains a linear correlation between longitudinal position and energy deviation after the arc. Distortions near the distribution edges are attributed to longitudinal space-charge effects. Additionally, the increase in uncorrelated energy spread is due to the ISR effect, while the change in distribution chirp is primarily driven by CSR effects and compression.
There is about 5.4 keV uncorrelated energy increase through the 90-degree arc.

\begin{figure}[!htb]
\centering
\includegraphics[width=6.0cm]{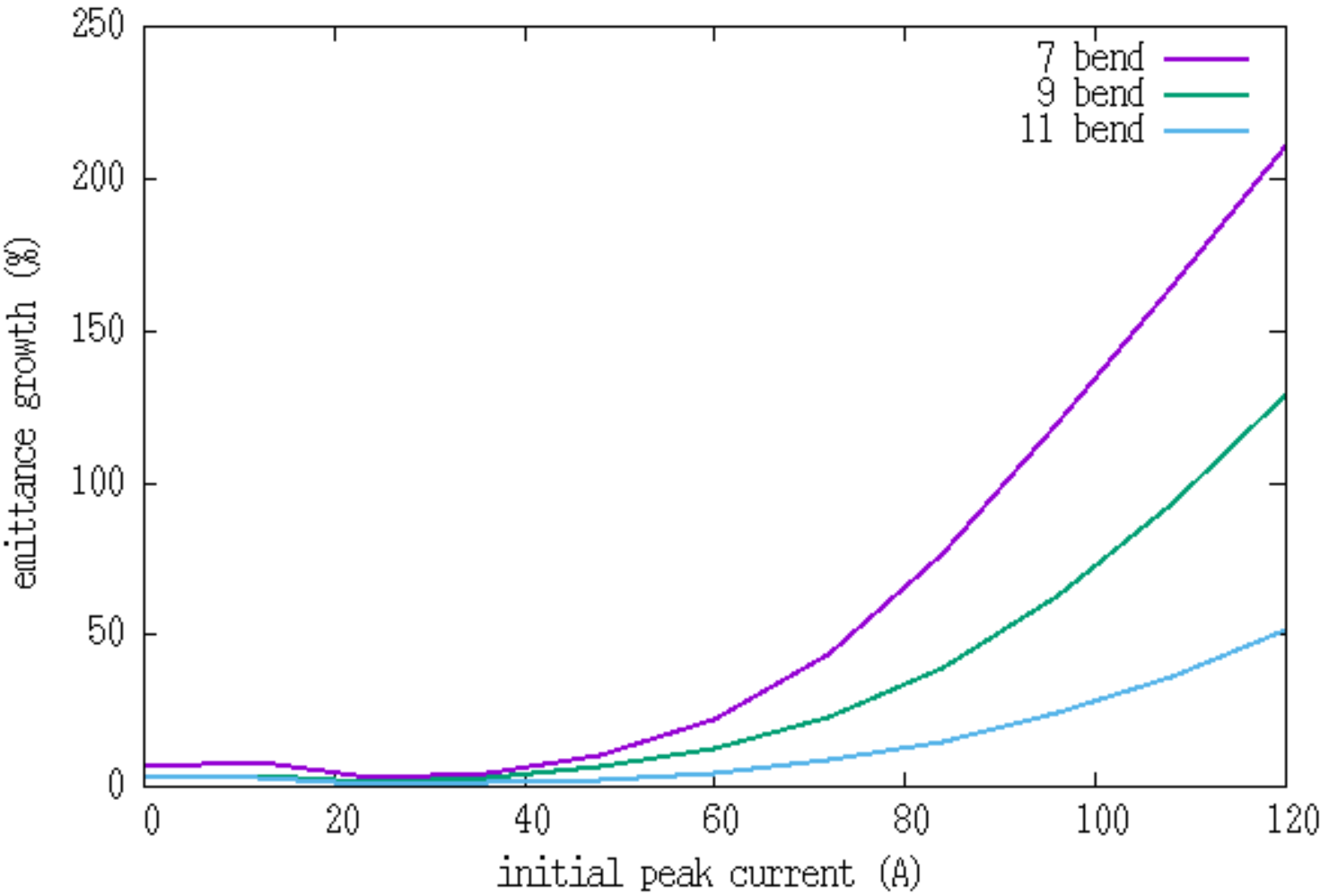}
\includegraphics[width=6.0cm]{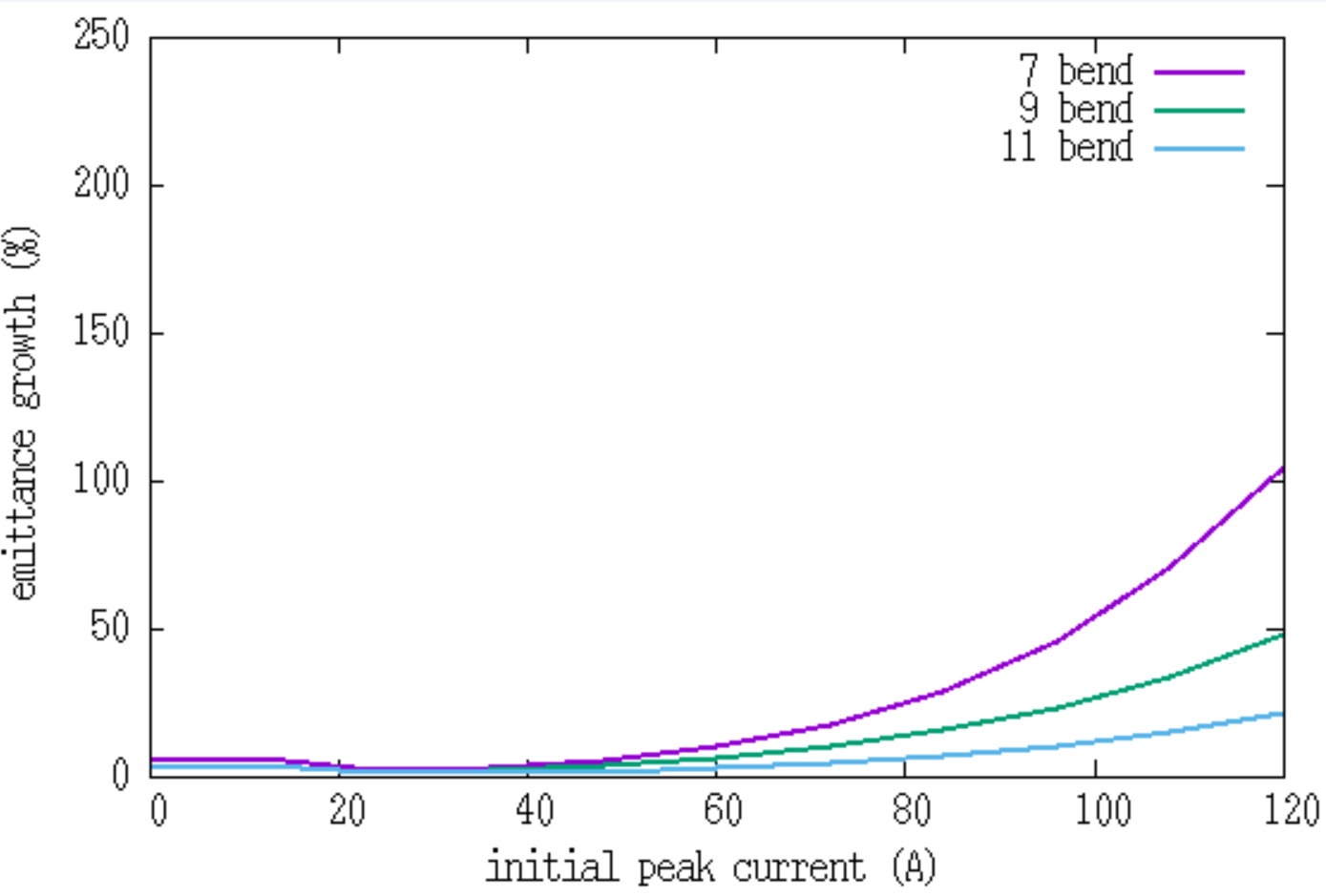}
\caption{Horizontal emittance growth through 7-bend, 9-bend, and 11-bend achromat arcs with an initial emittance of 0.5~$\mu$m (left) and 1~$\mu$m (right) at a beam energy of 1.2~GeV.}
\label{figemt12}
\end{figure}

In order to systematically evaluate the collective effects (primarily from CSR), we measured the final emittance growth after the arc for different initial peak currents with two initial emittances of 0.5~$\mu$m and 1~$\mu$m. We also studied the emittance growth using 7, 9, and 11 combined-function bends in a 90-degree arc, maintaining the same separation distance between consecutive bends. For a 1.2~GeV electron beam, the results are summarized in Fig.~\ref{figemt12}.

Across the range of initial peak currents studied, the emittance growth in the 7-bend arc is consistently higher than in the 9-bend and 11-bend arcs. Moreover, using more bending magnets in the arc results in smaller final emittance growth at the end of the arc. This may be attributed to the fact that a greater number of bending magnets leads to faster focusing and a smaller horizontal beam size within the bend, thereby reducing CSR-induced emittance growth. The final emittance growth through the 11-bend achromat arc can be kept below 10\% for initial peak currents below 70~A. To achieve a final emittance growth of less than 10\%, the initial peak current in the 9-bend arc should be less than 50~A, and less than 40~A in the 7-bend arc. With a larger initial emittance, the final emittance growth is reduced, though the basic trend remains unchanged. Using fewer bending magnets in a 90-degree arc reduces construction costs and the arc height (and thus the footprint), but it also increases the risk of larger emittance growth.

\begin{figure}[!htb]
\centering
\includegraphics[width=6.0cm]{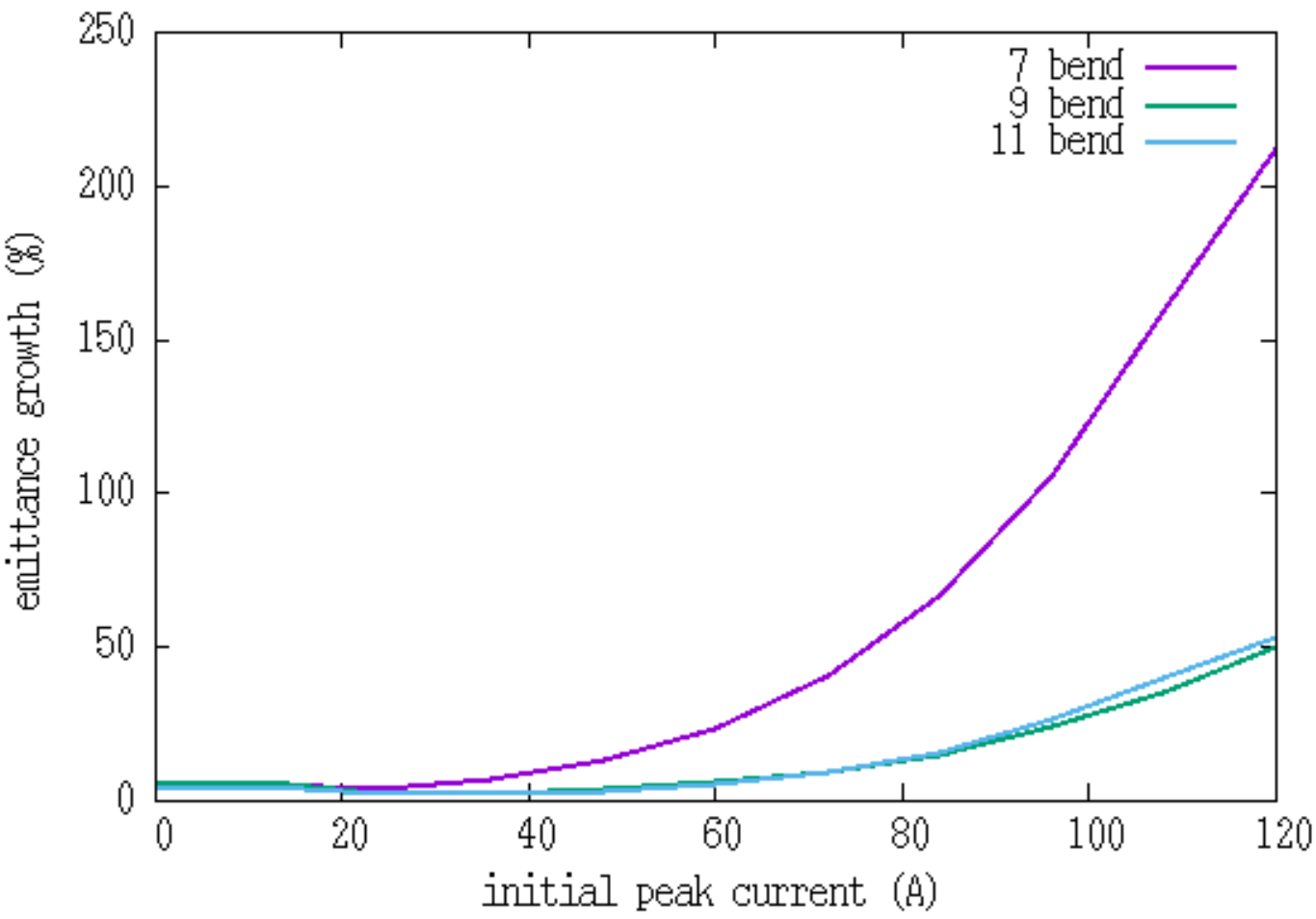}
\includegraphics[width=6.0cm]{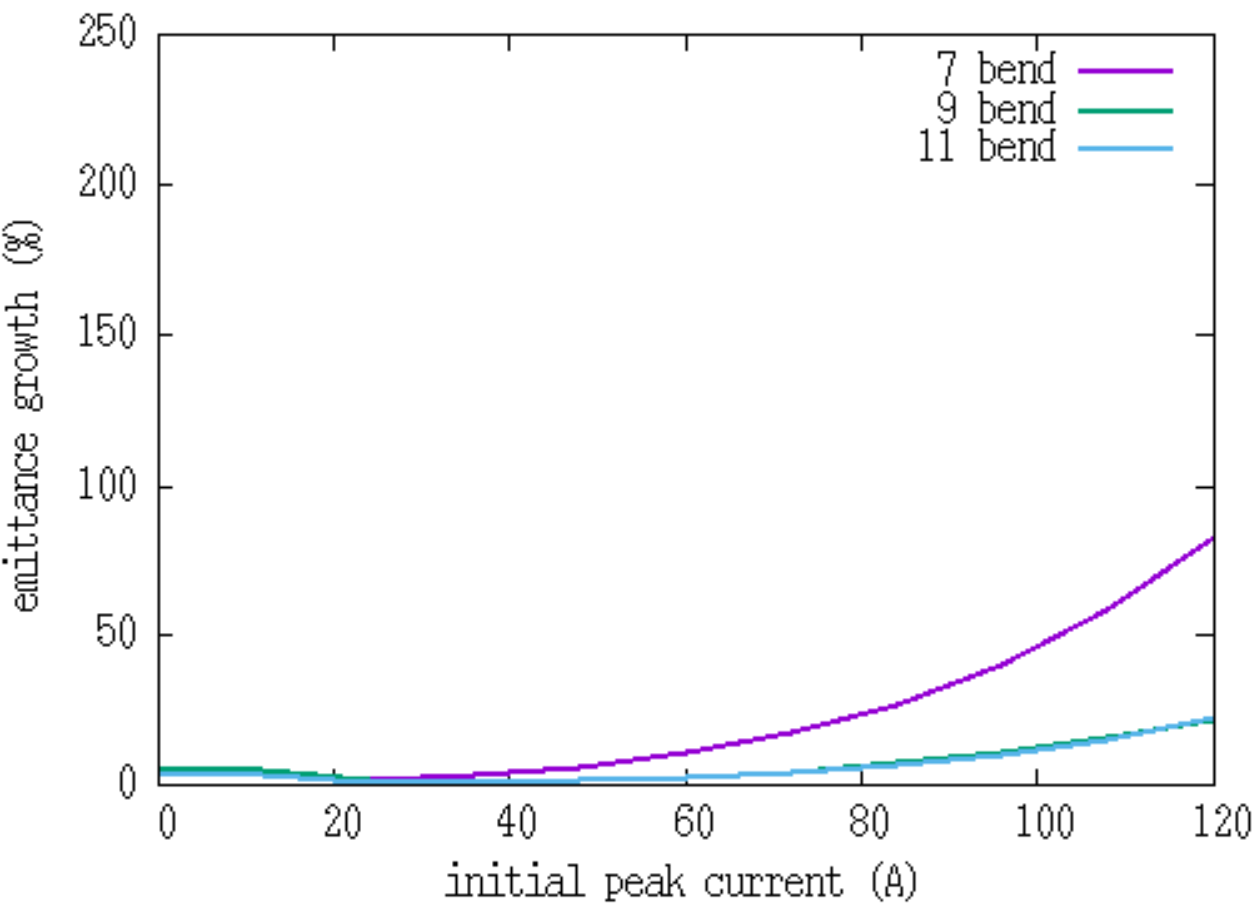}
\caption{Horizontal emittance growth through 7-bend, 9-bend, and 11-bend achromat arcs with an initial emittance of 0.5~$\mu$m (left) and 1~$\mu$m (right) at a beam energy of 1.5~GeV.}
\label{figemt15}
\end{figure}

We also measured the emittance growth as a function of initial peak current for an electron beam passing through 7-bend, 9-bend, and 11-bend arcs at a beam energy of 1.5~GeV. The simulation results are shown in Fig.~\ref{figemt15}. The emittance growth at this energy exhibits a similar pattern to the case at 1.2~GeV, except for a lower emittance growth through the 9-bend achromat arc. The emittance growth can be kept below 10\% for peak currents up to 70~A in the 9-bend and 11-bend arcs, and below 40~A for the 7-bend arc.

In all three bending magnet arcs, the vertical emittance growth at the end of the arc is small (less than 1\%).

\begin{figure}[!htb]
\centering
\includegraphics[width=6.0cm]{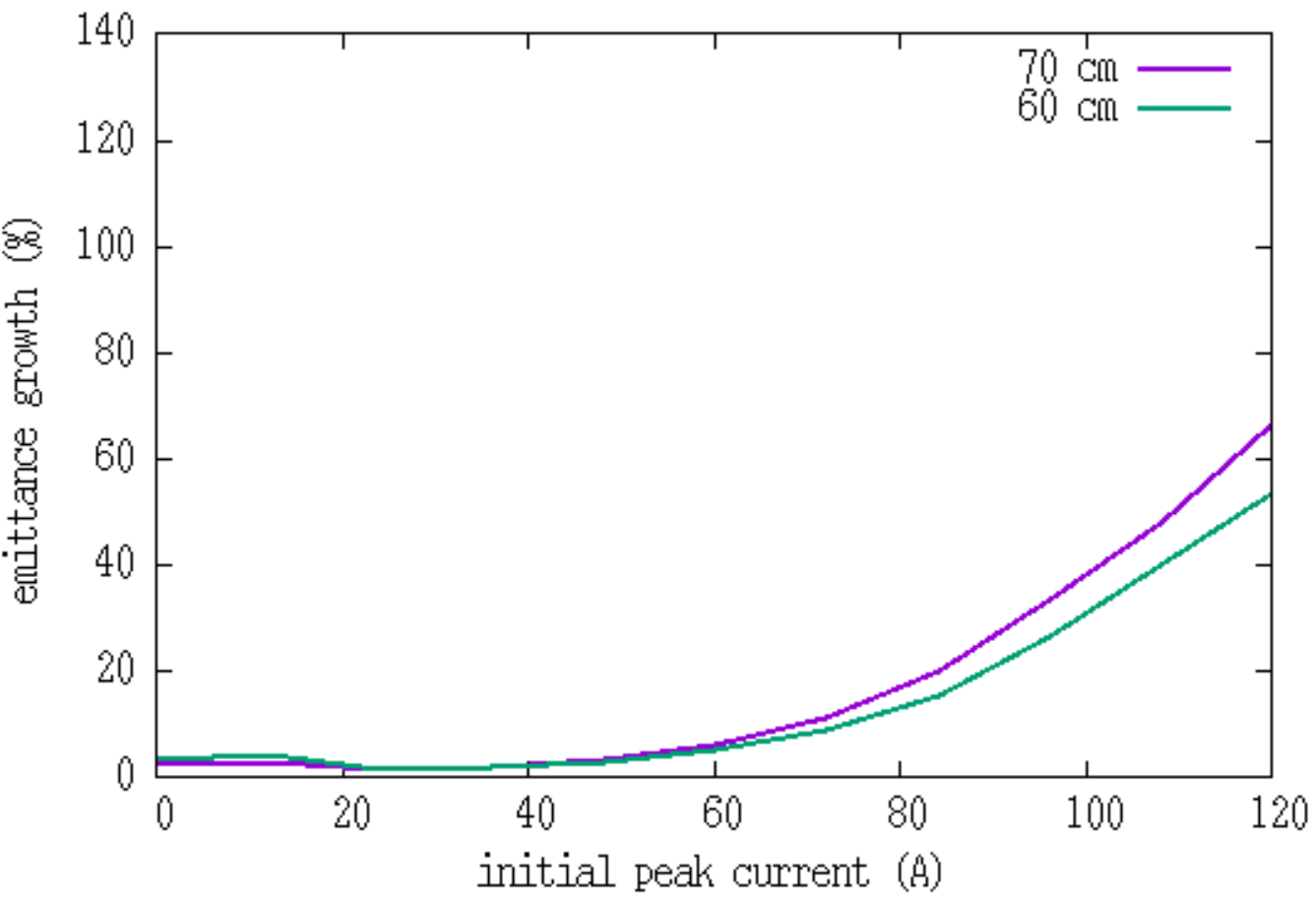}
\caption{Horizontal emittance growth through an 11-bend achromat arc with a 0.7~m separation between bending magnets, an initial emittance of 0.5~$\mu$m, and a beam energy of 1.5~GeV.}
\label{figemt07}
\end{figure}

In the above 90-degree arc, the separation between two bending magnets is assumed to be 0.6~m. This might be too tight for magnet hardware installation. We remeasured the emittance growth assuming a separation of 0.7~m between bending magnets in the 11-bend arc at 1.5~GeV. Figure~\ref{figemt07} shows the final emittance growth as a function of initial peak current with an initial emittance of 0.5~$\mu$m.

It is seen that there is slightly greater emittance growth across the initial current range with this larger separation. However, the final emittance growth can still be kept below 10\% for initial peak currents below 70~A. Increasing the separation between bending magnets would result in an increase in the arc height by approximately 0.5~m.

\begin{figure}[!htb]
\centering
\includegraphics[width=6.0cm]{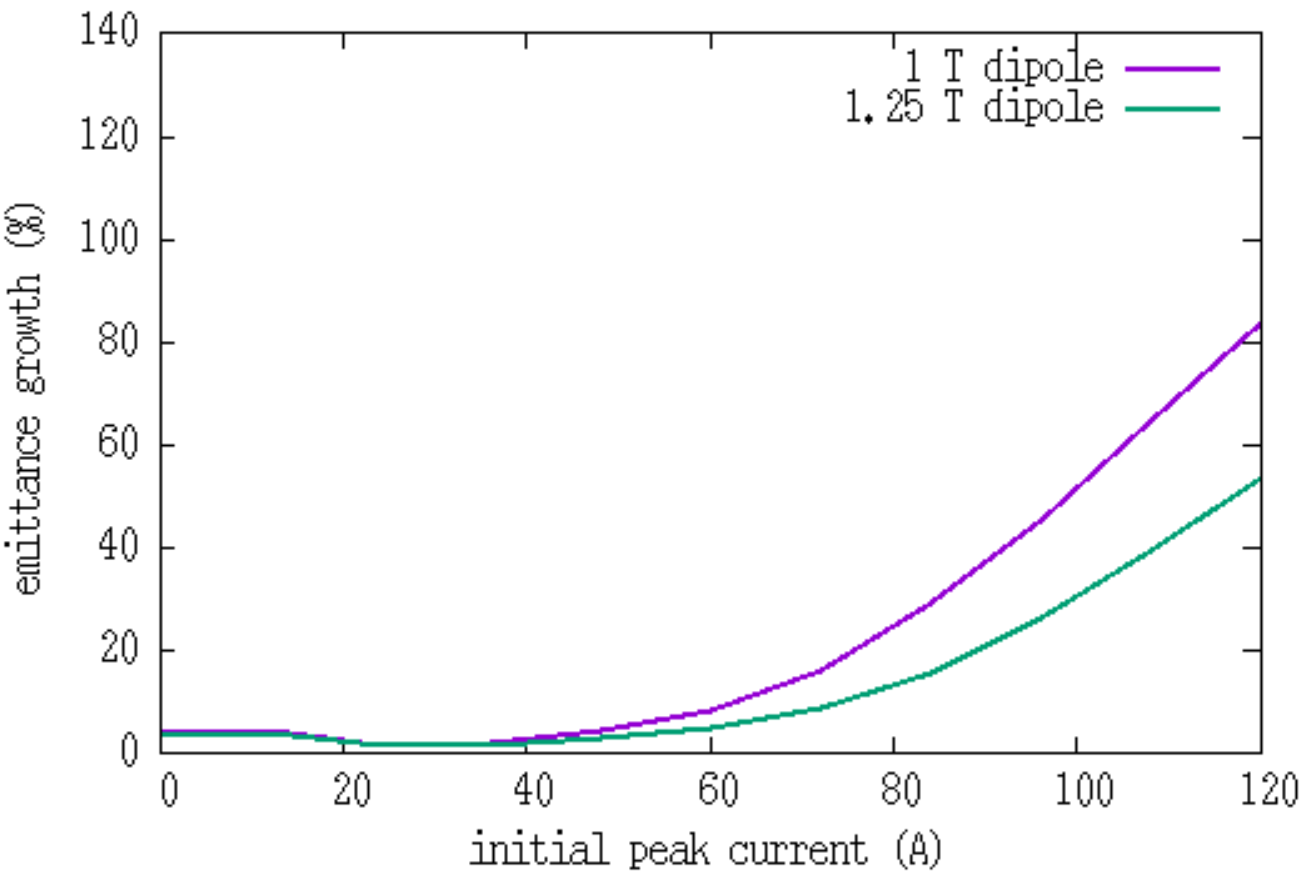}
\caption{Horizontal emittance growth through an 11-bend achromat arc with a 1~T magnetic field, an initial emittance of 0.5~$\mu$m, and a beam energy of 1.5~GeV.}
\label{figXemt1T}
\end{figure}

We also considered a case with a weaker magnetic field. In the above arcs, the magnetic field strength inside the bending magnet is assumed to be 1.25~T.
We investigated a weaker magnetic field case with a strength of 1~T inside the bending magnet. This results in a larger bending radius and a longer bending magnet length. Consequently, the arc height increases by approximately 1~m at 1.5~GeV. We remeasured the emittance growth assuming a 1~T magnetic field inside the bending magnets and a 0.6~m separation between them in the 11-bend arc at 1.5~GeV beam energy. Figure~\ref{figXemt1T} shows the final emittance growth as a function of initial peak current with an initial emittance of 0.5~$\mu$m.

There is slightly greater emittance growth across the initial current range with this weaker magnetic field inside the bends. Nevertheless, the final emittance growth can be kept below 10\% for most initial peak currents below 60~A.

There are a total of eleven 90-degree arcs in the proposed recirculating linac. With a 10\% emittance growth through each arc, this results in less than a factor of 3 growth in emittance from the injector. Based on the LCLS-II-HE injector design, an emittance of 0.1~$\mu$m can be achieved with a peak current of approximately 10~A~\cite{lcls2heinj0,lcls2heinj}. This suggests that the emittance through the recirculating linac (after the eleven arcs) and before the final compression can be less than 1~$\mu$m. Assuming the emittance growth after the final compression (by a factor of 10--20) is less than 100\%, the electron beam after the final bunch compression can achieve a peak current of 1~kA with an emittance of less than 2~$\mu$m.

\begin{figure}[!htb]
\centering
\includegraphics[width=6.0cm,height=4.5cm]{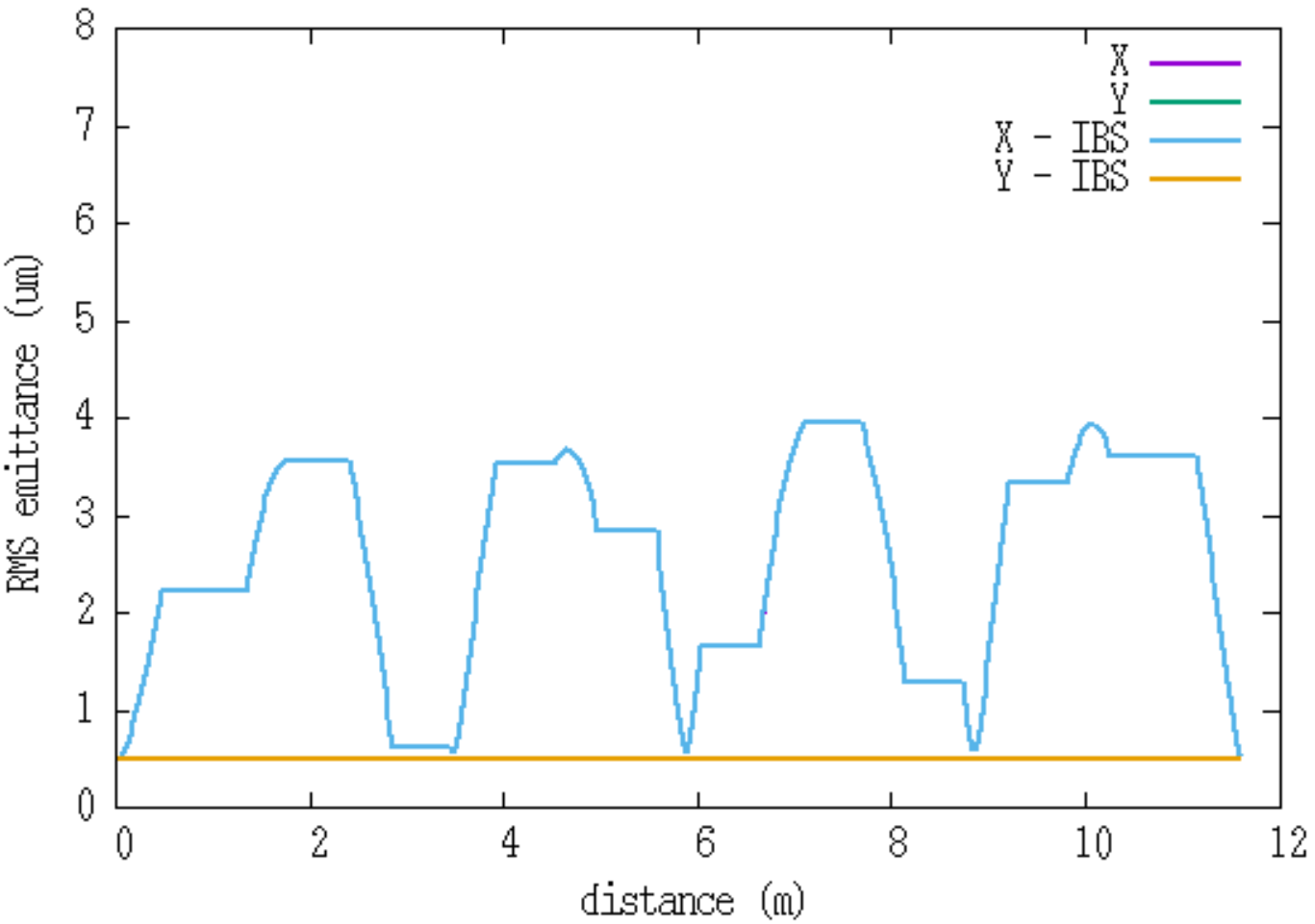}
\caption{Electron beam RMS emittance evolution through a 90-degree, 11-bend achromat arc with a beam energy of 1.2~GeV and an initial peak current of 60~A with and without including the IBS effect.}
\label{figibs}
\end{figure}

In the above simulations, we did not include the intrabeam scattering (IBS) effect from binary Coulomb collisions.
Subsequently, we used a Langevin approach to solve the Fokker-Planck-Landau equation self-consistently~\cite{qiangibs}. Figure~\ref{figibs} illustrates the transverse emittance evolution of a 1.2 GeV electron beam through a 90-degree arc using 11 bending magnets, comparing results with and without the IBS effect.
The evolution curves are nearly identical, indicating that the contribution of IBS to transverse emittance growth is negligible. This is likely due to the short distance (approximately 12 meters) through the arc.

\section{Conclusion and Discussion}

In this paper, we propose a compact accelerator for MHz high-repetition-rate extreme ultraviolet to soft X-ray free electron laser radiation facility. Based on a recirculating superconducting linear accelerator, the concept features a footprint of less than 100 meters and utilizes five cryomodules similar to those in LCLS-II-HE, each containing eight 1.3 GHz superconducting RF cavities.

We have studied the challenging ISR and CSR effects in the proposed compact accelerator using
both analytical estimates and detailed numerical simulations. Our simulations account for lattice nonlinearity, space-charge effects, ISR, and CSR.
We also evaluated the effect of intrabeam scattering through the 90-degree arc.
With an appropriate choice of peak beam current, incoherent synchrotron radiation and coherent synchrotron radiation in the eleven 90° achromatic arcs are not expected to be limiting factors. 
These do not induce significant horizontal emittance growth, and their impact on vertical and slice emittances remains minimal. Additionally, the growth of uncorrelated energy spread is also modest.

The final 100 pC electron beam energy is approximately 1.8 GeV, with an energy spread of less than 500 keV, a peak current exceeding 1 kA, and a projected normalized emittance below 2 mm·mrad.
This electron beam can be directed to a farm of undulators to generate GW-level 1 nm X-ray radiation via either Self-Amplified Spontaneous Emission (SASE) or a seeded FEL.
A portion of the 1 MHz electron beam can be diverted to a separate beamline and post-accelerated to nearly 10 GeV using high-gradient acceleration technologies, such as X-band or cryo-cooled C-band structures~\cite{cband} or wakefield accelerators~\cite{zholents}.
This high-energy beam can then be utilized to generate hard X-ray FEL radiation.

Several challenges remain in the development of this compact accelerator for XFEL facility, including the design of four merging sections for entering and exiting the arcs, longitudinal phase space control, microbunching instability mitigation, multi-bunch beam breakup instability suppression, and optimization of the final bunch compressor design.

The electron beam entering or exiting the superconducting linac has different energies during each pass. This energy variation complicates the use of a single quadrupole lattice to match the beam into the two or three multi-bend achromat (MBA) arcs. This challenge can be addressed by extending the length of either the entrance or exit matching sections of the MBA. In the proposed compact recirculating linac, there are currently four meters between the bending magnet and the superconducting linac module. We can reduce this distance to provide more space for separated energy matching in each arc.
Following the first bending magnet, beams with different energies enter energy-specific matching sections before proceeding to the periodic cells. These matching sections would be individually designed to accommodate the specific beam energies and ensure proper matching of the Twiss parameters to the periodic cells. A similar beam spreader
and combiner system was designed in a previous study~\cite{sasha2}.

In our conceptual layout, sextupole magnets have not yet been incorporated. However, in addition to the third harmonic cavities located upstream of the first 90-degree arc, sextupole magnets can be implemented to control the longitudinal phase space effectively and to reduce emittance growth~\cite{england}.

Microbunching instability represents a significant challenge in X-ray FEL accelerators, as it can substantially degrade the final electron beam quality~\cite{ubiexp,qiang17}. Several mitigation strategies are available, including installing a laser heater in the first straight section~\cite{LH} or incorporating low-beta quadrupole lattices to exploit intrabeam scattering for beam heating~\cite{qiangibs}. Additionally, controlled dispersion leakage from the arcs can be utilized to suppress microbunching instability~\cite{qiang13,li}.

Within the recirculating superconducting linac, the electron beam traverses the superconducting cavities multiple times. It is crucial to suppress beam breakup instability arising from higher-order modes (HOMs) excited when off-axis beams pass through the RF cavities. This instability can be mitigated through specialized optics designs or the implementation of HOM dampers~\cite{bbu1,tennant,bbu3}.

The design of the final bunch compressor may utilize an arc compression system with sufficiently large $R_{56}$ values~\cite{arclike0,arclike0b,arclike1,arclike2}.

In addition, future work includes the design of a passive dechirper and a beam switch yard~\cite{placidi}. Once these components are finalized, comprehensive start-to-end simulations will be conducted to evaluate the overall FEL performance, including all collective effects. Prior to construction, detailed simulations incorporating machine imperfections, alignment errors, and beam steering corrections should be performed to validate the design robustness.

\section*{ACKNOWLEDGEMENTS}
We would like to thank comments from Drs. C. Geddes, Z. Huang, A. Zholents... This work was supported by the U.S. 
Department of Energy (DOE), Office of Science, Office of 
High Energy Physics, under Contract No. DE-AC02- 
05CH11231. Computational resources were provided by 
the National Energy Research Scientific Computing Center 
(NERSC), which is supported by the DOE Office of 
Science under the same contract number.

	{%
	

} 
%
%


\end{document}